\documentclass[preprint2]{emulateapj}

\newcommand{\mr}{\mathrm}
\shorttitle{First {\it RHESSI} quiet Sun observations.}

\shortauthors{Hannah et al.}

\begin{document}

\title{First limits on the 3-200 keV X-ray spectrum \\of the quiet Sun using {\it RHESSI}}

\author{I. G. Hannah, G. J. Hurford , H. S. Hudson and R. P. Lin}
\affil{Space Sciences Laboratory,
University of California at Berkeley,
\\Berkeley, CA, 94720-7450, USA}
\author{K. van Bibber}
\affil{Physics and Advanced Technologies Directorate, \\Lawrence
Livermore National Laboratory, University of California,
\\Livermore, California, 94550, USA}

\email{hannah@ssl.berkeley.edu}
\email{hurford@ssl.berkeley.edu,hudson@ssl.berkeley.edu}
\email{rplin@ssl.berkeley.edu,vanbibber1@llnl.gov}

\begin{abstract}
We present the first results using the Reuven Ramaty High-Energy Solar
Spectroscopic Imager, {\it RHESSI}, to observe solar X-ray emission not
associated with active regions, sunspots or flares (the quiet Sun). Using a
newly developed chopping technique (fan-beam modulation) during
seven periods of offpointing between June 2005 to October 2006, we
obtained upper limits over 3-200~keV for the quietest times when the {\it
GOES12} 1-8~\AA~ flux fell below $10^{-8}$~Wm$^{-2}$. These values
are smaller than previous limits in the 17-120~keV range and extend
them to both lower and higher energies. The limit in 3-6~keV is
consistent with a coronal temperature $\leq 6$~MK. For quiet Sun periods
when the {\it GOES12} 1-8~\AA~ background flux was between
$10^{-8}$~Wm$^{-2}$ and $10^{-7}$~Wm$^{-2}$, the {\it RHESSI}
3-6~keV flux correlates to this as a power-law, with an index of $1.08
\pm 0.13$. The power-law correlation for microflares has a steeper index
of $1.29 \pm 0.06$. We also discuss the possibility of observing quiet Sun
X-rays due to solar axions and use the {\it RHESSI} quiet Sun limits to
estimate the axion-to-photon coupling constant for two different axion
emission scenarios.
\end{abstract}

\keywords{Sun: X-rays, gamma rays -- Sun: activity -- Sun: corona --
elementary particles}

\section{Introduction}

The X-ray spectrum of the Sun free of sunspots, active regions and flares
(the quiet Sun) is an important yet elusive measurement, despite interest
back to the earliest days of solar X-ray observations
\citep[e.g.,][]{neupert1969}. Such an observation would provide insight
into the nature of possible small-scale steady-state energisation
processes in the solar corona. For \emph{soft} X-rays (i.e., X-rays emitted
by thermal sources as free-free, free-bound continua or lines) the solar
corona is comparable  to other stars \citep[e.g.,][]{pevtsov2001}. The
stellar coronal emission consists of contributions from more than one
physical component with an emission measure distributed over
temperature. For \emph{hard} X-rays (usually characterised by an
arbitrary minimum photon energy instead of defined as non-thermal
bremsstrahlung) there is only one reported observation, that of
\citet{peterson1966}.

Observations of soft X-ray emission not associated with active regions,
for instance with {\it Yohkoh/SXT} \citep{harvey1992}, show X-ray bright
points that are weak compared to active region emission. They are
numerous, well-dispersed across the solar disk and are associated with
network boundaries. The presence of non-thermal electrons in these
events has been inferred from radio observations \citep{krucker1997} but
no hard X-ray emission was detected. This is because previous hard X-ray
imaging observations (Solar Maximum Mission{\it /HXIS}, {\it Hinotori},
and {\it Yohkoh/HXT}) were optimized to study flares, and were ill-suited
to observe weak sources distributed over large angular scales.

With small flares, i.e. microflares, the presence of non-thermal electrons
has been confirmed by microwave \citep{gary1997} and {\it RHESSI} hard
X-ray observations \citep{krucker2002}. Imaged {\it RHESSI} microflares
are always associated with active regions \citep{hannah2006agu}.
Although we might expect hard X-ray emission from below the current
limit of microflare observability, it is uncertain whether such a population
would exist in the absence of active regions. It is speculated that still
smaller non-thermal energy releases, such as ``nanoflares''
\citep{parker1988}, could produce globally distributed hard X-rays.

In addition to these processes, the interaction of cosmic rays in the solar
atmosphere could also generate weak diffuse X-rays from the quiet Sun
\citep{dolan-fazio1965,seckel1991}. Additional nuclear processes arising
from such cosmic ray interactions are likely to only produce minuscule
X-ray emission : for instance \emph{inner bremsstrahlung} from
$\beta$-decaying neutrons in the solar analog of the ``CRAND''
(Cosmic-Ray Albedo Neutron Decay) mechanism \citep{mackinnon07}, is
predicted to produce X-rays at a level far below that of the diffuse cosmic
background, which is $10^{-4}$ to
$10^{-8}~\mr{ph}~\mr{s}^{-1}~\mr{cm}^{-2}~\mr{keV}^{-1}$ from 3~keV
to 100~keV over a solar disk area.

Axions \citep{weinberg1978,wilczek1978} are hypothetical
weakly-interacting particles that could also produce an X-ray signature
from the Sun \citep{sikivie1983}. Nuclear reactions in stellar cores should
produce axions copiously; in the case of the Sun the average energy of
axions is 4.2~keV \citep{vanbibber1989}. These axions can convert
directly to X-ray photons in a perpendicular magnetic field
\citep{sikivie1983}, with the resulting photons having the same energy
and momentum as the incident axion. Ground-based experiments using
strong magnetic fields have tried to use this process to search for solar
axions \citep{zioutas2005}. The probability of this conversion is
proportional to the square of the product of the axion-photon coupling,
the distance travelled through a perpendicular magnetic field, and the
strength of this field \citep{sikivie1983}. This raises the possibility of
conversion in the corona \citep{carlson1996}. Attempting to observe such
a small flux would be difficult but would be more favourable during quiet
Sun periods when the conventional X-ray emission from the Sun is at a
minimum.

The Reuven Ramaty High-Energy Solar Spectroscopic Imager, {\it RHESSI}
\citep{lin2002}, has unprecedented sensitivity for 3-25~keV X-rays
because when its automated attenuators are ``out''  it can observe with
the full area its detectors. This was not possible for earlier instruments
which used fixed shielding to prevent excessive counting rates from soft
X-rays in flares. Normal {\it RHESSI} imaging is accomplished with a set of
nine bigrid rotating modulation collimators (RMCs) with resolutions
logarithmically spaced from 2.3 to 183 arcseconds. Each RMC
time-modulates sources whose size scale is smaller than their resolution.
Thus despite its sensitivity, {\it RHESSI} is not well suited to observe
weak  sources larger than $\sim3$~arcminutes. Most potential
mechanisms for quiet-Sun emission would be expected to be weak and
well-dispersed across the~32~arcminute solar disk.

For weak sources it is essential to distinguish counts due to solar photons
from counts due to terrestrial, cosmic or instrumental background. We
adopt an offpointing technique called \emph{fan-beam modulation}
\citep{hannah_fbm}, that provides a time-modulated, or ``chopped'',
signal of the solar disk, allowing us to distinguish distributed solar
emission from the background.

In this letter we detail the first analysis of periods of quiet Sun with {\it
RHESSI} using the fan-beam modulation technique (\S\ref{sec:method}).
We present the first limits of the quiet Sun X-ray spectrum and show how
this correlates with {\it GOES12} 1-8~\AA~flux and {\it RHESSI}
microflares in \S\ref{sec:data}. In \S\ref{sec:axions} we discuss the X-ray
emission due to solar axions and in \S\ref{sec:discuss} we discuss the
further work that can be achieved using with fan-beam modulation during
solar minimum conditions.

\section{Fan-beam Modulation Technique}\label{sec:method}
Instead of using the rapid time modulation associated with its RMCs,
\emph{fan-beam modulation} is based on a secondary modulation that
results from the finite thickness of {\it RHESSI's} collimator grids
\citep{hurford2002}. Fan-beam modulation depends upon the offpointing
angle, with a maximum effect when {\it RHESSI} is between $0.4^\circ$
and $1^\circ$ from offpointed Sun center \citep{hannah_fbm}. This
``envelope'' modulation peaks twice every rotation when the slits of the
grids are parallel to the line between {\it RHESSI} pointing and source
center, producing two transmission maxima per rotation.  For a period of
offpointing we bin (or ``stack'') the data in a chosen energy range
according to the roll angle of the spacecraft. These data are fitted with
the expected modulation and the resulting amplitude is corrected for the
predicted grid transmission efficiency. This technique works best in {\it
RHESSI's} RMC with the narrowest field of view, RMCs 1 to 6. In addition,
RMCs 2, 5 and 7 are not used in this analysis as they have the poorest
energy response.

This type of observation has been done for a total of 45~days during
seven periods (2005~July~19-26, 2005~October~18-28,
2006~January~12-17, 2006~February~1-7, 2006~August~3-8,
2006~September~26-29 and 2006~October~12-23), when the {\it
GOES12} 1-8~\AA~flux was around  $10^{-8}$~Wm$^{-2}$ and no active
regions or spots were on the disk. The Sun was very quiet during these
periods with the microwave emission (F10.7 levels) in the range
70-78~SFU and the {\it GOES12} 1-8~\AA~flux ``flat-lining'' below
$10^{-8}$~Wm$^{-2}$, the equivalent of an A1~class flare.

The data set was divided into five-minute time intervals, which are short
enough so that the radial offset, and hence grid transmission factor,
changes little. For the quiet Sun results presented in this paper we have
removed intervals with sharp time-series features (such as flares or
particle events). The analysis was further restricted to times when {\it
RHESSI} is at the lowest latitudes in its orbit, to minimize the terrestrial
background. From these four offpointing periods we have a total of 1,774
five-minute time intervals (over 147 hours of data), 1,071 (or
89.25~hours) of which occurred while the {\it GOES12} 1-8~\AA~flux was
below the A1 class of $10^{-8}$~Wm$^{-2}$. For each of these time
intervals, and over chosen energy bands, we have an amplitude of the
quiet Sun count rate corrected for the transmission through {\it RHESSI's}
grids. The resulting fitted amplitudes are then combined from different
time intervals to improve the signal to noise before conversion to a final
photon flux using the diagonal elements of the appropriate detector
response matrix \citep{smith2002}.

\begin{deluxetable}{ccc}
\tablecolumns{3} \tablewidth{0pc} \tablecaption{\label{tab:results}The
weighted mean, and its associated standard deviation, of quiet Sun
photon flux for periods when {\it GOES12} 1-8~\AA~flux
$<10^{-8}$~Wm$^{-2}$.}
 \tablehead{\colhead{Energy} &\colhead{Weighted Mean}
 &\colhead{$\sigma$}\\
 \colhead{keV} & \multicolumn{2}{c}
 {$\times 10^{-4} \mr{ph}~\mr{s}^{-1}~\mr{cm}^{-2}~\mr{keV}^{-1}$}}
 \startdata
3--6&  330.99 &$\pm 207.25$ \\
  6--12&  -5.24 &$\pm 8.46$ \\
  12--25&  -0.73 &$\pm 1.34$ \\
  25--50&  0.14&$\pm 0.63$ \\
  50--100& 0.74 &$\pm 0.54$ \\
  100--200&  -0.79& $\pm 0.42$\\
\enddata
\end{deluxetable}

\section{{\it RHESSI} Quiet Sun Spectrum}\label{sec:data}

\begin{figure}\centering
\plotone{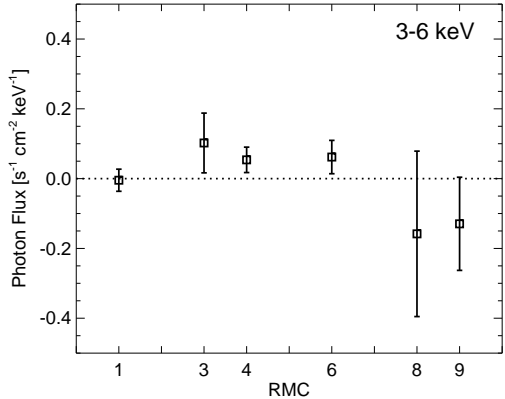} \caption{
The 3-6~keV flux observed in each of {\it RHESSI's} rotation modulation
collimators (RMC) averaged over the times when {\it GOES12}
1-8~\AA~flux was below the A1 class. The errors are larger with RMCs 8
and 9 since their FOV makes them less efficient for fan-beam modulation.
The dotted line indicates the zero flux level, the case if there
was no solar emission above the background.} \label{fig:fbmvsrmc}
\end{figure}

Figure \ref{fig:fbmvsrmc} shows the average 3-6 keV emission for six of
{\it RHESSI's} detectors. There is a small scatter between the detectors,
but consistent to within the errors. To calculate the average overall flux
for each energy band, we use the weighted mean of the value found in
RMCs 1, 3, 4 and 6, and give the error as the standard deviation in this
weighted mean. Table~\ref{tab:results} shows these values for energy
bands between 3 to 200~keV during the quietest times.

\begin{figure}\centering
\plotone{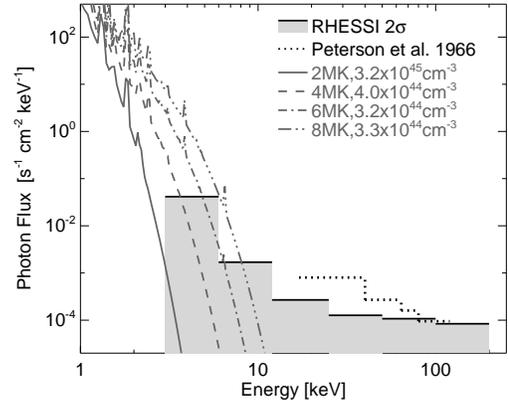} \caption{
The $2\sigma$ upper limits of the quiet Sun photon flux spectrum,
calculated using the {\it RHESSI} time intervals when the {\it GOES12} flux
in the 1-8~\AA~band fell below the A1 class, $10^{-8}$~Wm$^{-2}$. The
values shown are the 2$\sigma$ limits, from the standard deviation of
the weighted mean of the four RMCs. The \citet{peterson1966} limits are
quoted as having a ``95\% confidence''. The four thermal spectra are
CHIANTI models using an observation of the quiet corona with {\it
Yohkoh}/SXT \citep{pevtsov2001}, to constrain the possible temperature
and emission measures.}\label{fig:ph3200spec}
\end{figure}

None of the values given in Table~\ref{tab:results}  show a clear
statistical significance. Therefore we do not claim detection of a signal
from the quiet Sun but rather give conservative 2$\sigma$ upper limits to
the quiet Sun emission using the errors given in Table~\ref{tab:results},
shown in Figure~\ref{fig:ph3200spec}.  The dotted histogram shows the
data of \cite{peterson1966}, who made pioneering solar hard X-ray
observations with balloon-borne scintillation counters, in the 17~to
120~keV range. Our hard X-ray upper limits improve upon than these
results to about 75~keV and are above this energy. Our results also
extend the energy range into the previously unmeasured domains below
17~keV and above 120~keV.

Figure~\ref{fig:ph3200spec} also shows four CHIANTI thermal spectral
models for coronal emission \citep{dere1997,landi2006}. For each
assumed temperature the emission measure is constrained by {\it
Yohkoh}/SXT observations, which was sensitive to 1-2~keV X-rays, during
solar minimum \citep{pevtsov2001}. The {\it RHESSI} limit in 3-6~keV is
consistent with a quiet coronal temperature $\leq 6$~MK.

\begin{figure}\centering
\plotone{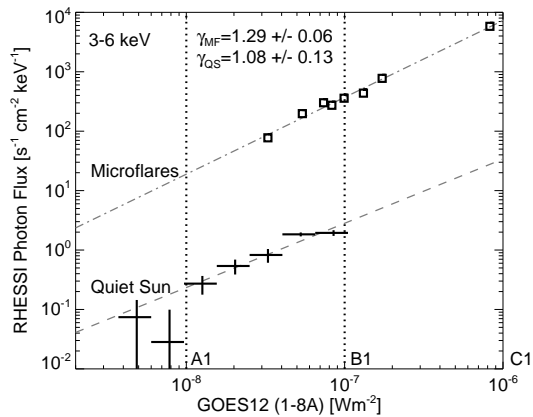} \caption{
The 3-6~keV photon flux from the fan-beam modulation technique for the
quiet Sun (broad crosses) and eight microflares during offpointing (square
data points) as compared to the corresponding 1-8~\AA~{\it GOES12}
flux. The {\it GOES12} flux for the microflares is background subtracted.
The fitted line for the quiet Sun points uses the five data points above {\it
GOES12} A1 class:  below this level the {\it GOES12} data digitises as it
reaches its sensitivity limit. }\label{fig:qsvsgs}
\end{figure}

We can gain a better understanding of possible quiet Sun X-ray emission
by using all 1,774 five-minute time intervals, not just the quietest
periods. These quiet Sun observations occurred over a range of {\it
GOES12} 1-8~\AA~ background fluxes up to $10^{-7}$~Wm$^{-2}$ an still
in the absence of active regions. By calculating the {\it RHESSI} flux in
consecutive subsets of {\it GOES12} background fluxes, we can plot the
correlation of {\it RHESSI} quiet Sun 3-6~keV flux to {\it GOES12}, shown
by the broad crosses in Figure~\ref{fig:qsvsgs}. The errors shown here are
the statistical ones found from the fit errors from each time interval
combined in quadrature. There is a clear power-law correlation between
the {\it GOES12} and {\it RHESSI} data, with an index of $\gamma=1.08
\pm 0.13$.

To put these observations in context we have also shown, as the square
data points in Figure~\ref{fig:qsvsgs}, the fluxes for eight microflares
that occurred during quiet Sun offpointing. The times of these microflares
were excluded from our quiet Sun 5 minute time intervals, since they are
a sign of activity. We used the fan-beam modulation technique for 16
seconds about the peak of these flares and plotted the flux against the
corresponding background subtracted {\it GOES12} flux. The {\it RHESSI}
flux from these microflares is around two orders of magnitude larger than
the quiet Sun values and there is again a power-law correlation between
{\it RHESSI} and {\it GOES12}. The microflare power-law correlation with
{\it GOES12} is slightly steeper ($1.29\pm0.06$) than that of the quiet
Sun.

\section{Solar Axions}\label{sec:axions}

Axions emitted from the burning core of the Sun may be converted to
X-rays by its own coronal magnetic field \citep{carlson1996}, thus
providing a detectable signal during periods of solar quiescence.  The
Sun's general field is constant and well-constrained during quiet periods,
and so it should be possible to derive a robust limit on the axion-photon
coupling $g_\mr{a\gamma\gamma}$, provided that other conventional
solar mechanisms can be convincingly excluded.

\citet{carlson1996} calculated whether such X-ray emission was
observable by assuming $g_\mr{a\gamma\gamma} =
10^{-10}~\mr{GeV}^{-1}$ and a dipole field scaled from a $10^{-4}$ T
polar field, predicting a flux of
$4\times10^{-2}~\mr{ph}~\mr{s}^{-1}~\mr{cm}^{-2}~\mr{keV}^{-1}$ over
3-6 keV. This is valid for the case of a sufficiently light axions, i.e.
$m_\mr{a}  < 1.8 \times 10^{-6}$ eV. This X-ray flux is comparable to the
value given in Table \ref{tab:results}. The $g_\mr{a\gamma\gamma}$
used in this calculation is similar to those cited by \citet{zioutas2005} in
a direct search for solar axions, and the best bounds from stellar
evolution, i.e. those from Horizontal Branch Stars
\citep{raffelt1996,raffelt2006}. This indicates that the {\it RHESSI} limit is
consistent with other approaches, albeit for smaller axion masses.

\citet{zioutas2004} published limits utilizing {\it RHESSI} data,
constraining massive Kaluza-Klein (KK) axions, which arise in certain
theories of large extra dimensions, in a scenario where KK axions are
emitted from the Sun in gravitationally bound orbits, and subsequently
undergo free-space decay $a \rightarrow \gamma \gamma$. Their value
of $g_\mr{a \gamma \gamma}$ however was derived from {\it RHESSI}
data taken during solar maximum and had not been corrected for
instrumental response. Using the flux estimate in Table \ref{tab:results},
we obtain an X-ray luminosity about two orders of magnitude smaller
than the value cited by \citet{zioutas2004}; and repeating their
calculation, we find a somewhat smaller limit to the axion-photon
coupling constant within this scenario of $g_\mr{a \gamma \gamma} \ll 6
\times 10^{-15}~\mr{GeV}^{-1}$.

\section{Discussion \& Conclusions}\label{sec:discuss}
We have established new upper limits on quiet-Sun emission in hard
X-rays when activity levels were below {\it GOES12}~A1 class. The most
natural explanation of such possible emission is that of unresolved
microflares. Figure~\ref{fig:qsvsgs} however, indicates that such
emission would require an ensemble of very small microflares, as the {\it
RHESSI} flux for resolved microflares is over two orders larger than that of
the quiet Sun. In this regime the energy release maybe due to a
mechanism physically different from (micro-) flares, such as the
speculated nanoflares \citep{parker1988}. These quiet Sun limits must be
interpreted in terms of our knowledge of the distribution function of flare
magnitudes, which follows a flat power law \citep{hudson1991}. The
comparisons in Figure~\ref{fig:qsvsgs} would then yield a normalisation
of the distribution law (e.g., the constant multiplier in the power-law
distribution). This will be calculated once the {\it RHESSI} microflare
distribution is known -- such a study is near completion
\citep{hannah2006agu}.

The work presented here represents the first use of the \emph{fan-beam
modulation technique} and shows that this method opens a new regime
of observations for {\it RHESSI}, namely weak sources larger than 3
arcminutes in size. Further use of this technique during the quieter times
of solar minimum should help improve the limits presented here.

The detection of the component of the solar X-ray flux due to axions
converting in the coronal magnetic field is especially challenging.
\citet{carlson1996} suggested that the stronger magnetic fields
associated with sunspots should permit tighter limits and are valid up to
higher values of the axion mass. However such limits would be subject to
greater uncertainties associated with the modeling of these magnetic
fields. Using {\it RHESSI} to search for this emission due to axions may be
more effective despite the large background from conventional emission
since we know the characteristic spatial scale (i.e. core size) on which
the emission is expected to occur \citep{vanbibber1989} and it should
vary in a distinctive manner as the sunspot moves across the solar disk.
Given {\it RHESSI's} low Earth orbit, it may be preferable to observe the
X-rays produced through the conversion of axions in Earth's nightside
magnetic field \citep{davoudiasl2005}. This method has the advantage
that the Earth blocks the competing solar X-ray flux.

A detailed presentation of {\it RHESSI} limits on solar axions from both
daytime and nighttime observations of the Sun will be the subject of a
subsequent paper.

\section{Acknowledgements}
NASA supported this work under grant NAG5-98033. Work at Lawrence
Livermore National Laboratory was supported under Department of
Energy Contract W-7405-Eng-048. We thank many people involved in the
{\it RHESSI} program for inspiration and assistance with software,
calibration, and spectral fitting issues, especially Brian Dennis, Richard
Schwartz and David Smith.


\end{document}